\documentclass[twoside,twocolumn,american,showpacs]{revtex4-1}
\usepackage[T1]{fontenc}
\usepackage[latin9]{inputenc}
\usepackage{amsmath}
\usepackage{amssymb}
\usepackage{graphicx}

\usepackage{color}
\makeatletter
\@ifundefined{textcolor}{}
{%
 \definecolor{BLACK}{gray}{0}
 \definecolor{WHITE}{gray}{1}
 \definecolor{RED}{rgb}{1,0,0}
 \definecolor{GREEN}{rgb}{0,1,0}
 \definecolor{BLUE}{rgb}{0,0,1}
 \definecolor{CYAN}{cmyk}{1,0,0,0}
 \definecolor{MAGENTA}{cmyk}{0,1,0,0}
 \definecolor{YELLOW}{cmyk}{0,0,1,0}
}

\@ifundefined{date}{}{\date{}}
\makeatother

\usepackage{babel}
\begin{document}

\title{Crafting networks to achieve, or not achieve, chaotic states.}

\author{Sarah De Nigris }

\email{denigris.sarah@gmail.com}

\affiliation{Departement of Mathematics and Namur Center for Complex Systems-naXys, University of Namur, 8 rempart de la Vierge 
5000 Namur, Belgium}

\author{Xavier Leoncini}

\email{Xavier.Leoncini@cpt.univ-mrs.fr}

\affiliation{Aix Marseille Universit\'e, Universit\'e de Toulon, CNRS, CPT UMR 7332,
13288 Marseille, France}
\begin{abstract}
The influence of networks topology on collective properties of dynamical
systems defined upon it is studied in the thermodynamic limit. A network
model construction scheme is proposed where the number of links and the
average eccentricity are controlled.
This is done by rewiring links of a regular one dimensional chain
according to a probability $p$ within a specific range $r$, that
can depend on the number of vertices $N$. We compute the thermodynamical
behavior of a system defined on the network, the $XY-$rotors model,
and monitor how it is affected by the topological changes. We identify
the network {effective} dimension $d$ as a crucial parameter: topologies with
$d<2$ exhibit no phase transitions while ones with $d>2$ display
a second order phase transition. Topologies with $d=2$ exhibit states
characterized by infinite susceptibility and macroscopic chaotic/turbulent
dynamical behavior. These features are also captured by $d$ in the
finite size context.
\end{abstract}

\maketitle
Networks are ubiquitous in the reality surrounding us and indeed the
network perspective for systems of interacting agents has been a real
paradigm shift in various realms ranging from physics to biology,
sociology and economics \cite{dorogotsev_evolution,Newman_book,havlin_book,Barrat_Book}.
One pivotal feature shared by many existing networks, like the World
Wide Web \cite{Broder2000,Barabasi_diameter,Barabasi2000} or social
networks \cite{Milgram1969}, is the so called 'small-world' property:
two nodes are separated by a short path consisting in just few edges
thanks to the presence of long-range connections, the \emph{shortcuts},
in the network. Since this property often arises in a self-organized
fashion, it could seem natural at first to infer that those shortcuts
favor the flow of information and more easily lead to collective states,
like if a kind of evolutionary principle is at play. But are indeed
those long-range links always beneficial to enhance global coherence?
A striking example of this dilemma can be the brain: from one side
it displays the small-world property \cite{Review_brain} but, at
the same time, there are evidences of\emph{ chaotic} response in living
neural systems \cite{freeman_neurons}. In contrast, small-world topologies
can be a fertile substrate to enhance transport phenomena as navigation
\cite{Kleinberg2000} and, more recently, it has also been shown that
the overall conductance of a network is advantaged by the introduction
of long-range links \cite{Oliveira2014}. It hence appears highly
non trivial, when dealing with interacting agents upon a network,
to ask oneself what kind of collective behavior they can possibly
display since a chaotic response can arise along with a coherent one
due to the presence of long-range links.

This work inscribes itself in this frame: we provide here a mean to
construct a class of networks in which the addition of long-range
links can give rise to a whole range of dynamical and statistical
behaviors and, in particular, it also entails a chaotic state of infinite
susceptibility, similar to the one encountered in \cite{deNigris2013,deNigrisPrE_2013}.
Moreover our network model is crafted to embed real networks characteristics
but via minimal assumptions so to ensure a certain form of simplicity.
As we will display in the following, we related the different behaviors
to the network \emph{dimension}, which changes according to the injection
of long-range links.

\begin{figure}
\includegraphics[width=4cm,height=4cm]{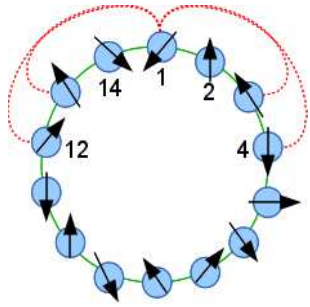}
\caption{(color online) Practical network construction for $N=14$, $\gamma=1.2$
thus $k\propto\left\lfloor N^{0.2}\right\rfloor =2$ and $r=\left\lfloor \sqrt{N}\right\rfloor =3$.
The starting configuration is the solid (green) line since we have
just two links per spin and the dotted (red) links are the possible
rewiring.\label{fig:Practical example}}
\end{figure}

\begin{figure}
\includegraphics[width=8.5cm,height=6.5cm]{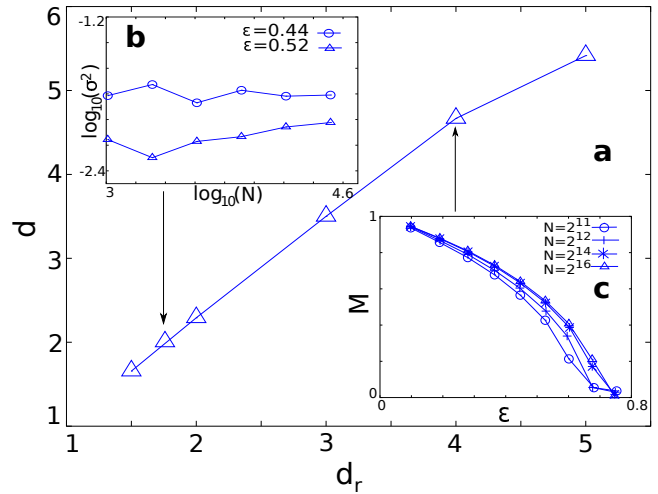}\caption{(color online) (a) Dimension of a completely rewired network $(p=1)$
with $N=2^{14}$ and $r\propto N^{\delta}$. The horizontal axis is
the parametrization in Eq.~(\ref{eq:random dimension}) which
gives, with our choice of $r$, $d_{r}=1/(1-\delta)$.\label{fig:dimension N}
(b) Scaling of the magnetization variance $\sigma^{2}(N)$ with the
system size $N$ for $d=2$. (c) Phase transition of the magnetization
$M(\varepsilon),\,\varepsilon=E/N$ for $d>4$. The error bars are
within the dots size.}
\end{figure}

In our model, networks are built starting from
a {k-regular network with periodic boundary conditions and degree}
$k\propto N^{\gamma-1}(1\leq\gamma\leq2)$ (where $N$ is the number of vertices), which constitutes
the backbone. {Practically, the nodes are laid on a one dimensional ring and each of them interacts with its $k$ closest neighbors (see Fig.~\ref{fig:Practical example}}).
{Therefore our starting configuration is completely symmetrical and invariant under rotations.} In this work, we set $\gamma$ close to $1$, $\gamma=1.2$,
in order to have a few links per vertex (for instance we get $k=12$
for $N=2^{14}$). {  This choice for the present work is meant to recover sparseness \cite{Barrat_Book}, which is a common feature in many real world networks;
nevertheless the influence of the $k$ parameter by itself was explored in \cite{deNigris2013}}.
We then proceed to a construction
similar to the Watts-Strogatz one for Small World networks \cite{watts_strogatz1998SW}:
we rewire each link with probability $p$ but, differently from \cite{watts_strogatz1998SW},
we impose to rewire it within a range $r$ (Fig.~\ref{fig:Practical example}).
Therefore with our parameters $(\gamma,p,r)$ we put three ingredients
meant to mimic concrete systems: first the condition of \emph{sparseness}
through $\gamma$, i.e. a very low vertex degree compared to the system
size \cite{Barrat_Book}, second we introduced the concept of \emph{interaction
range} constraining the links to be at most of a fixed length $r$
and last we inject \emph{randomness }in the structure so to have a
non uniform degree.
 Hence, from one side, the range parameter $r$ mimics the fact that
in many natural and artificial systems interactions can occur only
within a certain neighborhood and on the other side the probability
$p$ ensures the presence of randomness in the link distribution,
so that all the length scales occur. The range concept is reminiscent
of the Kleinberg model \cite{Kleinberg_model} but, in our case, the
choice of $r$ entails a sharp cutoff in the distribution of the accessible
link lengths and, moreover, the probability $p$ to rewire a link
within the range $r$ is uniform. {Before proceeding we would like to stress
that the key parameter of interest for the present work is the range: indeed in 
two previous works \cite{deNigrisPrE_2013,deNigris2013} we investigated respectively the impact of the $k$
parameter on k-regular networks and the interaction between $k$ and $p$ for Small-World networks.
Now, with the range constraint, we practically enforce} a control
on the dimension: heuristically we can forecast that if we choose
for instance $r\propto\sqrt{N}$, the more links are rewired (i.e.
for high $p$), the more the network will be shaped like a bi-dimensional
object, because we have in some sense crafted from the initial ring
a $\sqrt{N}\times\sqrt{N}$ lattice. To give a more quantitative counterpart
to this view, we define the dimension $d$ similarly to the dimension
on Euclidean lattices: for the latter, it holds a power law relation
between the volume and a characteristic length $V\propto r^{d}$,
the exponent $d$ being the dimension. Then in our context of networks,
we have to consider a specific length scale. Here we settled for the
average of the vertices eccentricity $ec(i)$, i.e. the longest path
$\ell_{i,j}$~~$i\neq j$ attached to each vertex $i$. Thus we define
our characteristic length $\ell$ as: 
\begin{equation}
\ell=\frac{1}{N}\sum_{i}ec(i).\label{eq:eccentricity}
\end{equation}
Hence if we consider its scaling with the network
volume (size) $N$, we obtain a definition of dimension: 
\begin{equation}
d=\frac{\log N}{\log\ell},\label{eq:dimension}
\end{equation}
The definition in Eq.~(\ref{eq:dimension}) recovers in the $N\rightarrow\infty$
limit the one already proposed in \cite{Havlin1991,Baglietto2013,newman1999scaling}
in which they consider the power law scaling of the average path length
$\ell_{av}$ with the network size $N$, while we take in account
in Eq.~(\ref{eq:eccentricity}) the average vertex eccentricity $\ell_{ec}$.
These two quantities are indeed related since $\ell_{ec}\sim2\ell_{av}$
and this assumption was also numerically tested. It is hence evident
that the difference between the two dimension definitions is a term
vanishing logarithmically with the size $N$, thus proving their equivalence
in the $N\rightarrow\infty$ limit. However, in the range of system
sizes used in our simulations, the definition in Eq.~(\ref{eq:eccentricity})
was the more suitable choice to grasp the dimension since the aforementioned
difference is still important enough to introduce a small shift in
the dimension value.
In Eq.~(\ref{eq:dimension}), it is straightforward to see that
the dimension of the completely rewired ($p=1$) configuration is
intrinsically related to the range $r$: indeed for $p=1$, we have that $\ell\sim N/r$ since each node very probably possesses a link rewired 
at a distance $r$. Therefore, if $\ell\sim N/r$, we have that Eq.~(\ref{eq:dimension}) becomes
\begin{equation}
d_{r}=\frac{\log N}{\log N-\log r}.\label{eq:random dimension}
\end{equation}
{In what follows we shall use
the dimension $d_r$ given by Eq.~(\ref{eq:random dimension}) as our control parameter: in practice $d_{r}$ corresponds to a re-parametrization 
of the range which we will consider to be of the type $r\sim N^\delta$ with $N\gg1$ and $\delta>\gamma-1$.}
If we take our previous example of $r=\sqrt{N}$, we obtain that the
corresponding network with $p=1$ has indeed $d_{r}=2$ and, in Fig.~\ref{fig:dimension N}(a),
we display how the measured network dimension for $p=1$ follows Eq.~(\ref{eq:random dimension})
so that, fixing the range $r(N)$, we can control the resulting dimension
once we have rewired all the links, independently from the size.

Having thus an operative and general way to set and quantify the dimension,
we used our network model to investigate the thermodynamic response
of a dynamical system defined upon these networks and test the influence
of the dimension $d$ in Eq.~(\ref{eq:dimension}). With this goal in mind, we consider $N$ $XY-$rotors
\cite{book_introduction_renorm,book_condensed_matter}, whose dynamics
is described by an angle $\theta_{i}(t)$ and its canonically associated
momentum $p_{i}(t)$.
{We shall show that the rewiring of a few links, beyond altering significantly the network structure,
can also entail different collective behaviors: in particular, we shall investigate if, like on regular lattices, we have  
a spontaneous symmetry breaking for $d>2$, which is absent when $d<2$. This brings some analogies to the extension of 
the Mermin-Wagner theorem on inhomogeneous structures \cite{cassi1992,cassi1996} in which the critical parameter to discriminate 
between different regimes is the spectral  dimension \cite{cassi1999,cassi2000,cassi_review}, 
therefore opening an interesting thread of research.
Moreover we shall focus on $d=2$, or $r\sim\sqrt{N}$, to see if a chaotic state emerges, 
displaying some similarities woth the one observed in the regular structure discussed in 
\cite{deNigris2013}. Returning back to the $XY$-rotors,} 
each rotor $i$ is located on a network vertex
and its interactions are provided by the set $V_{i}$ of vertices
attached to it via the links. The Hamiltonian of the system  reads: 
\begin{equation}
H=\sum_{i=1}^{N}\frac{p_{i}^{2}}{2}+\frac{J}{2\left\langle k\right\rangle }\sum_{i\in V_{i}}(1-\cos(\theta_{i}-\theta_{j})),\label{eq:hamiltonian}
\end{equation}
where $J>0$, $\left\langle k\right\rangle $ is the average degree and $V_{i}=\{j\neq i |~\exists ~e_{i,j}\in~E\}$, $E$ being the ensemble of edges.
The dynamics of the network is given by the two Hamilton equations:
\begin{equation}
\begin{cases}
\dot{\theta_{i}} & =p_{i}\\
\dot{p}_{i} & =-\frac{J}{\left\langle k\right\rangle }\sum_{j\in V_{i}}\sin(\theta_{i}-\theta_{j})
\end{cases},\label{eq:dynamics}
\end{equation}
We run molecular dynamics (MD) simulations of the isolated system
in Eqs.~(\ref{eq:dynamics}), starting with Gaussian initial conditions
for $\left\{ \theta_{i},p_{i}\right\} $. The simulations are performed
integrating the dynamic equations in Eqs.~(\ref{eq:dynamics}) with
the fifth order optimal symplectic integrator, described in \cite{mclachlan1999accuracy},
with a time step of $\Delta t=0.05$. Such an integrating scheme allows
us to check the correctness of the numerical integration since we
verified at each time step that the conserved quantities of the system,
the energy $E=H$ and the total momentum $P=\sum_{i}p_{i}/N$, are
effectively constant. The total momentum $P$ is set at $0$ as initial
condition without loss of generality. In order to grasp the amount
of coherence in the system, we define a magnetization $M=\left|\mathrm{\mathbf{M}}\right|$
as order parameter: 
\begin{equation}
\mathrm{\mathbf{M}}=\frac{1}{N}\sum_{i}\left(\cos\theta_{i},\sin\theta_{i}\right)\label{eq:magnetizatio}
\end{equation}
and, once the system has reached a stationary state, we measure $\overline{M}$,
where the bar stands for the temporal mean. Thus, in the stationary
state, if $\overline{M}\sim1$, all the rotors point in the same direction,
whereas if $\overline{M}\sim0$ there is not a preferred direction.
Practically, once the network topology and the size $N$ are fixed,
we monitor the average magnetization $\overline{M(\varepsilon,N)}$ for each energy $\varepsilon=E/N$
in the physical range. We perform the temporal mean on the second
half of the simulation, after checking that the magnetization has
reached a stationary state, when it is reached (i.e. not in the case
of the chaotic state). The simulations time is typically of order
$Tf=10^{4}-10^{5}$. In the insets of Fig.~(\ref{fig:dimension N})
we display the dynamical response of the $XY$ model to different
dimensions: we chose $r$ so to have $d=2$ for $r\propto\sqrt{N}$
and $d>4$ for $r\propto N^{3/4}$. For the latter, in Fig.~(\ref{fig:dimension N})c
the magnetization displays a second order phase transition, seeming
to occur at $\varepsilon=E/N\sim0.75$, in the same fashion of the
Hamiltonian Mean Field (HMF) model \cite{Campa09}.
\begin{figure*}[!ht]
\begin{centering}
\includegraphics[width=8cm,height=6cm]{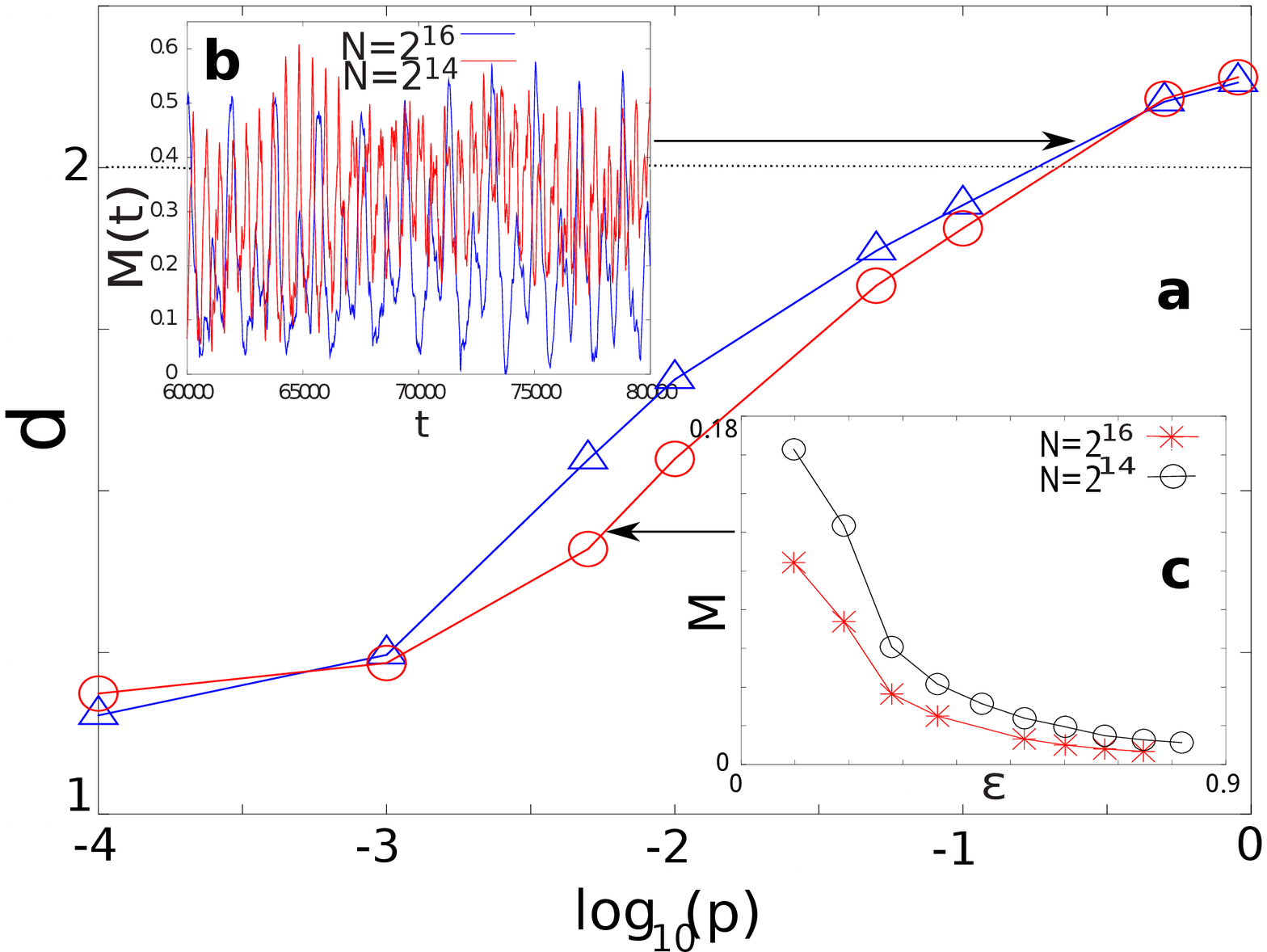} \includegraphics[width=8cm]{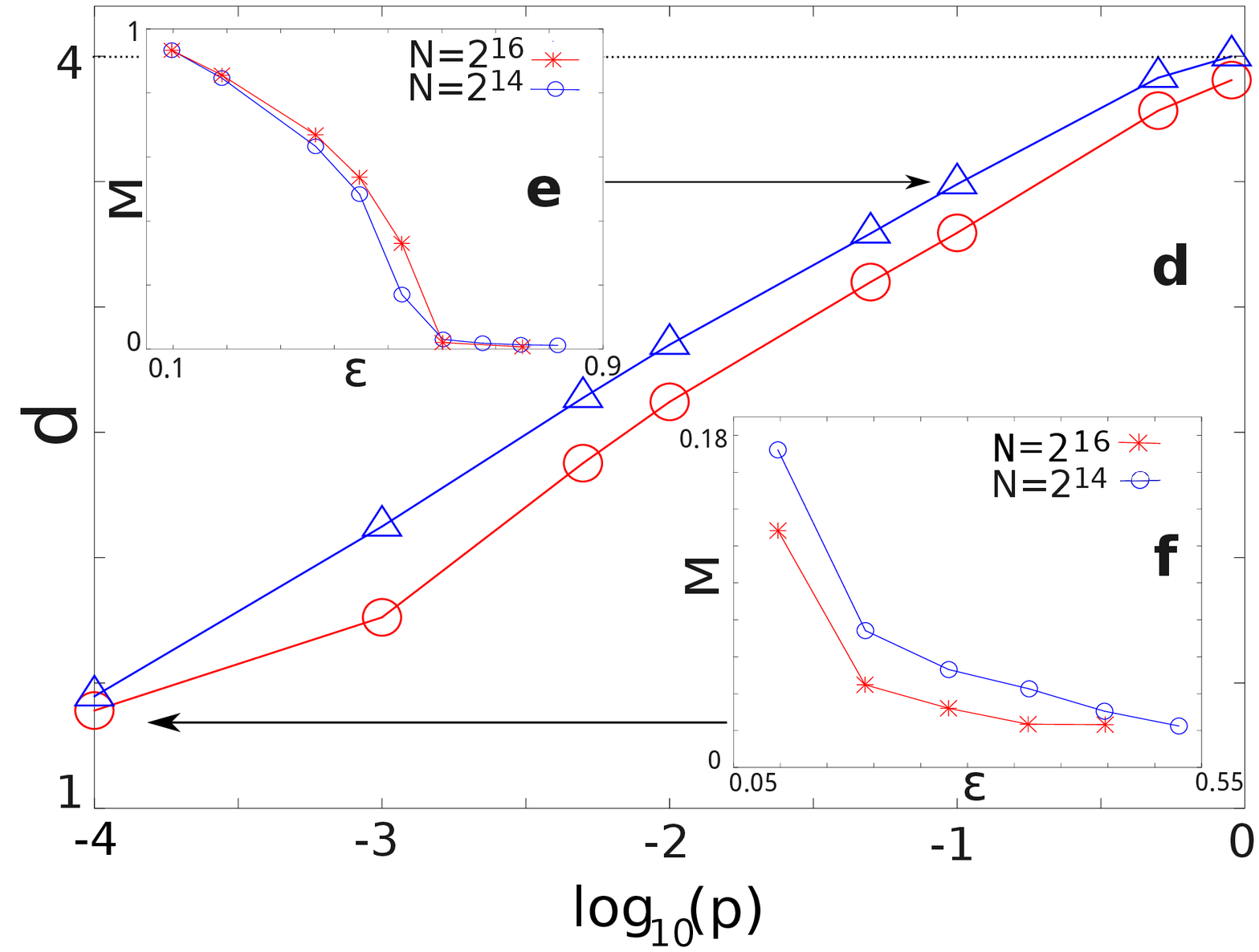}
\par\end{centering}

\caption{(color online) Dimension and its influence on global coherence. The
relation between the dimension $d$ and the fraction of links rewired,
given by $p$, for (a) $r=\sqrt{N}$ and (d) $r=N^{3/4}$ for two network
sizes, $N=2^{14}$(dots) and $N=2^{16}$(triangles). In (a) the dimension
shifts from 1 to 2, whereas in (d) the increased $r$ drags the dimension
up to $4$. In the insets we display the corresponding thermodynamical
response: in (b) for a network with $d=2$ the magnetization shows
a chaotic behavior at $\varepsilon=0.350(1)$ while in (c) and (f)
the quasi-unidimensional network does not sustain any long-range order,
entailing the vanishing of the magnetization for every energy. Finally,
for $d\simeq3$, (e) shows a second order phase transition at $\varepsilon_{c}=0.6$.
For the magnetization equilibrium values (c,e,f) and the dimension
(a,d), the error bars are within the dots size.}

\label{fig:Dimension-and-its} 
\end{figure*}
It is noteworthy
that, for the $XY$ model, the dimension $4$ is the one at which
mean field theory starts to apply and, indeed, the phase transition
displayed in Fig.~\ref{fig:dimension N}(c) for $d>4$ seems to confirm
this picture. For the case with $d=2$ we observed a state similar
to the one described in \cite{deNigrisPrE_2013,deNigris2013}: the
magnetization, for low energy densities $\varepsilon=E/N$, is affected
by important fluctuations like if the order parameter was oscillating
between the mean field value and zero. Moreover this regime does not
reach the equilibrium on the timescales considered: its persistence
was checked for simulation times $Tf\sim10^{6}$, i.e. $10$ times
longer that in previous cases and nevertheless it was not possible
to observe its relaxing. To give further insights on this chaotic
state arising in the network with $d=2$, we looked at the magnetization
variance $\sigma^{2}=\overline{(M-\overline{M})^{2}}$, where the
bar stands again for the temporal mean, in order to give a quantitative
measure of this regime. As shown in Fig.~\ref{fig:dimension N}(b)
the variance is unaffected by the size: this flat profile is in striking
contrast with the variance's canonical scaling $\sigma^{2}\propto1/N$,
leading to vanishing fluctuations in the thermodynamic limit. On the contrary, if we take into account
the definition of the magnetic susceptibility 
\begin{equation}
\chi\sim\lim_{N\rightarrow\infty}N\sigma^{2},\label{eq:chi}
\end{equation}
we have that this regime shall be characterized by an infinite susceptibility
in the thermodynamic limit.{  The peculiar nature of this regime is also highlighted by its persistence in an energy range. Indeed
in the usual $XY$ Kosterlitz-Thouless transition  the divergence of the susceptibility
occurs at the phase transition point \cite{KT73}, while  these ``turbulent'' states exist in a whole interval energies up to the critical one.
In fact these states are somewhat reminiscent of the observed \emph{quasi-stationary states} (QSS) 
occurring in the Hamiltonian Mean Field model or more generally in systems with long range interactions 
\cite{ettoumi2011linear,Ettoumi2013,chavanis2008out,van2010stationary,Levin2014}. Nevertheless, as mentioned, we do not observe any relaxation in contrast with 
the QSS's behaviour.}

Our model brings interesting perspectives for finite size systems
as well: as a first observation, we should note that our construction
procedure, like the Watts-Strogatz algorithm for Small World networks
\cite{watts_strogatz1998SW}, induces on average $N_{R}=Nkp\propto N^{\gamma}p$
rewired links. Hence the fraction of long-range connections increases
with the size (in the present study very slightly because of our choice
$\gamma=1.2$). We thus argue the existence of a non trivial interplay
between $p$, $r$ and $N$, so that it is possible, like for Small
World networks, to tune $p$ in order to change the measured dimension
for a given size $N$. In some sense, $d$ can turn out to be, for
a finite size system, a measure of an \emph{effective dimension} produced
by the fraction of rewired links. To test our hypothesis, we consider
$N=2^{14}$ and $N=2^{16}$ and in Fig.~\ref{fig:Dimension-and-its}(a)-(d)
we show how the progressive introduction of long-range links in the
network drags the dimension to $d=2$ for $r=\sqrt{N}$ (Fig.~\ref{fig:Dimension-and-its}(a))
and to $d=4$ for $r=N^{3/4}$(Fig.~\ref{fig:Dimension-and-its}(d)).
Indeed the shift between the two dimension curves mirrors the effect
of the two sizes and it is more pronounced for the largest range $r=N^{3/4}$.
Therefore a natural question arises: does the dynamical behavior relate
to this ``finite size'' dimension? Similarly to what we did in Fig.~\ref{fig:dimension N},
we analyzed the dynamical response of the $XY-$model and Fig.~\ref{fig:Dimension-and-its},
we display our results for $r\sim\sqrt{N}$ and $r\sim N^{3/4}$.
To guide our investigation, we can use Fig.~\ref{fig:Dimension-and-its}(a)-(d)
as a map to locate the parameter zones characterized by different
dimensions. Focusing first on $r\sim\sqrt{N}$ , we chose the probabilities
so to have either a network with $d=2$, $p=0.1$ and $p=0.3$,
or an quasi one-dimensional one, $p=0.005$. In Fig.~\ref{fig:Dimension-and-its}(b)
we show that indeed these networks generate a chaotic state similar
to the one in Fig.~\ref{fig:dimension N}(a) and described in \cite{deNigris2013,deNigrisPrE_2013}:
the heavy oscillations of the magnetization do not relax even for
long time simulations and their amplitude (i.e. the variance) is unaffected
by the size increase. This peculiar state, appearing for low energy
densities, seems again intrinsically related to the dimensionality
since the two aforementioned probabilities values entail $d\sim2$,
as displayed by Fig.~\ref{fig:Dimension-and-its}(a). Moreover we
considered several sizes to investigate the impact of the size increase
and, again, there is no significant difference between, for instance,  $N=2^{14}$
and $N=2^{16}$ in the fluctuations amplitude. On the other hand, it is
noteworthy to observe a signature of the different sizes in the oscillations
period, which is significantly slower in the $N=2^{16}$ case. This
effect, entangling system size and timescales, can be reminiscent with the lifetime of QSSs \cite{ettoumi2011linear,Ettoumi2013,chavanis2008out,van2010stationary,Levin2014}
of the HMF model, which is the mean field version of the $XY$ model.{ Moreover, the collective oscillation itself
recalls a very similar oscillating behavior observed in the HMF case \cite{morita2006collective} 
{or in the $\alpha-$HMF \cite{van2010stationary} for QSSs.
In this latter case of QSSs, this feature was used to perform ''Poincaré sections'' \cite{Bachelard08,van2010stationary}}. 
Nevertheless, we would like to stress that, both in \cite{deNigris2013} and in the present case, the root of the oscillating state is a 
\emph{topological} condition on the network and not a dynamical one, as the choice of a particular initial condition. Furthermore, as another
point of difference, we were not able to observe the eventual relaxation of those states so far.
Anyway, those analogies, like the aforementioned one on the phase transition, and those differences both point} to
very interesting research perspectives to shed light on the connection between these two systems.
Continuing in our analysis, for $p=0.005$ which gives $d\lesssim1.7$
(Fig.~\ref{fig:Dimension-and-its}(a)), the magnetization vanishes
for all the energies, so to confirm the crucial role played by the
crossover to the two dimensional configuration. Taking now into account
the case $r\sim N^{3/4}$, we show in Fig.~\ref{fig:Dimension-and-its}(e)
that the system undergoes a second order phase transition, as it happens
in Fig.~\ref{fig:dimension N}(c) when $d>2$. In Fig.~\ref{fig:Dimension-and-its}(e) the probability 
is set at $p=0.1$, which entails $d\sim 3$ for the sizes considered. On the other hand,
the short-range regime is at play for lower probabilities in Fig.~\ref{fig:Dimension-and-its}(f)
where we display the vanishing of the order parameter for $d\leq1.5$. 
In conclusion, we have provided a way to construct a class of networks
whose dimension $d$ is controllable via the range parameter $r$.
We have shown how this dimension, in the thermodynamic  limit,
is related to different collective states of the $XY$ model upon
those networks: for $d>2$ the system displays a second order phase
transition which becomes very similar to the one of the HMF model
for $d>4$, while for $d=2$ a regime characterized by an infinite
susceptibility is at play. Beyond the analysis in the thermodynamic limit, we
also interpreted the dimension $d$ in the case of finite size systems:
in this frame $d$ is a function of $(N,r,p)$ so that we can ``adjust''
the probability of rewiring $p$ to obtain the desired \emph{effective
dimension}.
{Considering the evidences we have displayed, we may argue that, for general networks, the considered dimension can be a key
topological characteristic that in the end governs the final collective behavior of large coupled
systems. Moreover we believe that the peculiar case of networks with $d=2$, for which the chaotic collective state emerges,
could lead to many interesting applications: for instance, the infinite susceptibility could be used to amplify signals,
or a better understanding of the dynamics could prove useful in the context of modeling and studying turbulent behaviors in 
an isolated system. On a closing note, the condition $d>2$ to have a collective behaviour, which is entangled with 
having a range of interaction $r>O(\sqrt N)$, bears a strong resemblance to the necessary condition for synchronization
of Kuramoto oscillators, as shown in \cite{mori2010}} and this latter analogy could point to the intrinsic importance of this 
topological feature over the details of the dynamic imposed on the network.

\section*{Acknowledgements}

The authors are grateful to a referee for having pointed Refs.~\cite{cassi1999,cassi1996,cassi2000,cassi1992}.
X. L. is partially supported by the FET project Multiplex 317532.

\bibliographystyle{apsrev4-1}

%

\end{document}